\documentclass[aps,prl,twocolumn,floats,showpacs]{revtex4}

\usepackage{epsfig}
\usepackage{color}
\usepackage{bm}
\usepackage{latexsym}

\begin{document}
\newcommand{\hide}[1]{}
\newcommand{\tbox}[1]{\mbox{\tiny #1}}
\newcommand{\half}{\mbox{\small $\frac{1}{2}$}}
\newcommand{\sinc}{\mbox{sinc}}
\newcommand{\const}{\mbox{const}}
\newcommand{\trc}{\mbox{trace}}
\newcommand{\intt}{\int\!\!\!\!\int }
\newcommand{\ointt}{\int\!\!\!\!\int\!\!\!\!\!\circ\ }
\newcommand{\eexp}{\mbox{e}^}
\newcommand{\bra}{\left\langle}
\newcommand{\ket}{\right\rangle}
\newcommand{\EPS} {\mbox{\LARGE $\epsilon$}}
\newcommand{\ar}{\mathsf r}
\newcommand{\im}{\mbox{Im}}
\newcommand{\re}{\mbox{Re}}
\newcommand{\bmsf}[1]{\bm{\mathsf{#1}}}
\newcommand{\mpg}[2][1.0\hsize]{\begin{minipage}[b]{#1}{#2}\end{minipage}}

\title{Diluted banded random matrices: Scaling behavior of eigenfunction and spectral properties}

\author{J. A. M\'endez-Berm\'udez,$^1$ Guilherme Ferraz de Arruda,$^{2,3}$ 
Francisco A. Rodrigues,$^2$ and Yamir Moreno$^{3,4,5}$}

\affiliation{
$^1$Instituto de F\'{\i}sica, Benem\'erita Universidad Aut\'onoma de Puebla, 
Apartado Postal J-48, Puebla 72570, Mexico \\
$^2$Departamento de Matem\'{a}tica Aplicada e Estat\'{i}stica, Instituto de 
Ci\^{e}ncias Matem\'{a}ticas e de Computa\c{c}\~{a}o, Universidade de S\~{a}o 
Paulo - Campus de S\~{a}o Carlos, Caixa Postal 668, 13560-970 S\~{a}o Carlos, SP, Brazil \\
$^3$Institute for Biocomputation and Physics of Complex Systems (BIFI), University 
of Zaragoza, Zaragoza 50009, Spain \\
$^4$Department of Theoretical Physics, University of Zaragoza, Zaragoza 50009,
Spain \\
$^5$Complex Networks and Systems Lagrange Lab, Institute for Scientific 
Interchange, Turin, Italy
}

\date{\today}

\begin{abstract}
We demonstrate that the normalised localization length $\beta$ of the eigenfunctions 
of diluted (sparse) banded random matrices follows the scaling law $\beta=x^*/(1+x^*)$. 
The scaling parameter of the model is defined as $x^*\propto(b_{\mbox{\tiny eff}}^2/N)^\delta$, 
where $b_{\mbox{\tiny eff}}$ is the average number of non-zero elements per
matrix row, $N$ is the matrix size, and $\delta\sim 1$.
Additionally, we show that $x^*$ also scales the spectral properties of the model 
(up to certain sparsity) characterized by the spacing distribution of eigenvalues.
\end{abstract}


\maketitle

{\it Introduction and outlook}.-- 
Random matrix (RM) models serve to describe statistical properties of complex 
systems and related processes: From the original Gaussian ensembles of Wigner 
and Dyson~\cite{Metha,RMT} (which reproduce the statistics of energy levels of complex 
nuclei, quantized chaotic systems, disordered systems, random networks, etc.) to very 
recent and more elaborated ensembles e.g.~relevant to the problem of many-body 
localization~\cite{KKCA}.

Even at the early years of RM modeling, Wigner himself realized the need to refine
the generic Gaussian ensembles in order to incorporate properties of realistic 
physical systems. In this respect he introduced the so-called Wigner-banded RM 
model~\cite{W55,W57} (see also Refs.~\cite{RMT,WFL91,FLW91,FGIM93,CCGI93,CCGI96,F97,W00,W01}), 
a model including a bandwidth and an increasing diagonal. In particular
the bandwidth, which quantifies the range of interactions, has been the main ingredient 
of other RM models proposed to deal with explicit applications: 
As examples we can mention 
the power-law banded RM model~\cite{MFDQS96,EM08} (used to simulate the 
Anderson metal-insulator transition), 
the banded random matrix (BRM) model~\cite{CMI90,EE90,FM91,CIM91,BMP91,FM92,MPK92,MF93,FM93,FM94,I95,MF96,CGM97,
KPI98,KIP99, P01,W02} (introduced to emulate quasi-one-dimensional disordered wires),
the embedded ensembles~\cite{MF75,BW03,K14} (which take into account the many-body 
interactions in complex nuclei and many-body systems), system-specific banded Hamiltonian 
RM models~\cite{CK01,CH00} (where the bandwidth of the Hamiltonian matrix can be 
obtained by means of semiclassical arguments~\cite{FP86,FLW91}), 
among many 
others~\cite{RMT,S94,FM95a,FM95b,S97,DPS02,KK02,S09,S10,CRBD16,MFRM16,FCIC96,MRV16}.

On the other hand, there exist several works dealing with diluted RM models, see for example 
Refs.~\cite{RB88,FM91b,MF91,E92,JMR01,KSV04,K08,S09b,SC012,EKYY13,EKYY12,MAMRP15}.
However, we know just a few RM models including, in addition to sparsity, an effective 
bandwidth: i.e.~the Wigner-banded RM model with sparsity~\cite{FCIC96}, 
diluted power-law RM models~\cite{CRBD16,MRV16}, 
and a diluted block-banded RM model~\cite{MFRM16}.

Thus, motivated by the ample interest on banded RM models and the recent attention
on diluted versions of them~\cite{CRBD16,MFRM16,MRV16} in this paper we study scaling 
properties of a diluted version of the BRM model. In particular we demonstrate that both
eigenfunction and spectral properties scale with a parameter that relates the model 
attributes (matrix size, bandwidth, and sparsity) in a highly non-trivial way. 

{\it Model definition and statement of the problem}.--
The BRM ensemble is defined 
as the set of $N\times N$ real symmetric matrices whose entries are independent Gaussian 
random variables with zero mean and variance $1+\delta_{i,j}$ if $|i-j|<b$ and zero
otherwise. Hence, the bandwidth $b$ is the number of nonzero elements in the first matrix 
row which equals 1 for diagonal, 2 for tridiagonal, and $N$ for matrices of the GOE. 
There are several numerical and theoretical studies available 
on this model, see for example 
Refs.~\cite{CMI90,EE90,FM91,CIM91,BMP91,FM92,MPK92,MF93,FM93,FM94,I95,MF96,CGM97,
KPI98,KIP99, P01,W02}. 
In particular, outstandingly, it has been found~\cite{CMI90,EE90,I95,FM91} 
that the eigenfunction properties of the BRM model, characterized by the {\it scaled localization 
length} $\beta$ (see Eq.~(\ref{beta}) below), are {\it universal} for the fixed ratio 
\begin{equation}
X = b^2/N \ .
\label{X}
\end{equation}
More specifically, it was numerically and theoretically shown that the scaling function
\begin{equation}
\beta = \frac{\Gamma X}{1+\Gamma X} \ ,
\label{betascaling0}
\end{equation}
with $\Gamma\sim 1$, holds for the eigenfunctions of the BRM model, see also 
Refs.~\cite{FM92,MF93,FM93,FM94}. 
It is relevant 
to mention that scaling (\ref{betascaling0}) was also shown to
be valid, when the scaling parameter $X$ is properly defined, for the kicked-rotator 
model~\cite{CGIS90,I90,I95} (a quantum-chaotic system characterized by a random-like 
banded Hamiltonian matrix), the one-dimensional Anderson model, and the Lloyd 
model~\cite{CGIFM92}.

We define the {\it diluted} BRM (dBRM) model by including sparsity, characterized by the
parameter $\alpha$, in the BRM model as follows: Starting with the BRM model we
randomly set off-diagonal matrix elements to zero such that the sparsity is defined as
the fraction of the $N(b-1)/2$ independent non-vanishing off-diagonal matrix elements. 
According to this definition, a diagonal random matrix is obtained
for $\alpha=0$, whereas the BRM model is recovered when $\alpha=1$.

Therefore, inspired by scaling studies of the BRM 
model~\cite{CMI90,EE90,CIM91,FM91,FM92,MF93,I95,CGM97,KIP99},
here  we propose the study of eigenfunction and spectral properties of the dBRM 
model as a function of the parameter
\begin{equation}
x = b_{\mbox{\tiny eff}}^2/N \ , \qquad b_{\mbox{\tiny eff}} \equiv \alpha b \ ,
\label{x}
\end{equation}
where the {\it effective bandwidth} of the dBRM model $b_{\mbox{\tiny eff}}$ is, in analogy
to the bandwidth $b$ of the BRM model, the {\it average} number of nonzero elements per 
matrix row.

{\it Eigenfunction properties}.--
A commonly accepted tool to characterize quantitatively the complexity of the 
eigenfunctions of random matrices (and of Hamiltonians corresponding to disordered 
and quantized chaotic systems) is the information or Shannon entropy $S$. This measures 
provides the number of principal components of an eigenfunction in a given basis. 
The Shannon entropy, which for the eigenfunction $\Psi^l$ is given as
\begin{equation}
\label{S}
S = -\sum_{n=1}^N (\Psi^m_n)^2 \ln (\Psi^m_n)^2 \ ,
\end{equation}
allows to compute the so called entropic eigenfunction localization length, 
see e.g.~\cite{I90},
\begin{equation}
\label{lH}
\ell_N = N \exp\left[ -\left( S_{\tbox{GOE}} - \bra S \ket \right)\right] \ ,
\end{equation}
where $S_{\tbox{GOE}}\approx\ln(N/2.07)$, which is used here as a reference, 
is the entropy of a random eigenfunction with Gaussian distributed amplitudes 
(i.e.,~an eigenfunction of the GOE). With this definition for $S$ when $\alpha=0$ or $b=1$, 
since the eigenfunctions of the dBRM model have only one non-vanishing component 
with magnitude equal to one, $\bra S \ket=0$ and $\ell_N\approx 2.07$. On the other 
hand, when $\alpha=0$ and $b=N$ we recover the GOE and $\bra S \ket=S_{\tbox{GOE}}$; 
so, the {\it fully chaotic} eigenfunctions extend over the $N$ available basis states
and $\ell_N\approx N$.

Here, as well as in BRM model studies, we look for the scaling properties of the 
eigenfunctions of the dBRM model through the {\it scaled localization length}
\begin{equation}
\beta = \ell_N/N \ ,
\label{beta}
\end{equation}
which can take values in the range $(0,1]$.

In the following we use exact numerical diagonalization to obtain the eigenfunctions 
$\Psi^m$ ($m=1\ldots N$) of large ensembles of dBRMs characterized by the  
parameters $N$, $b$, and $\alpha$. 
We perform the average $\bra S \ket$ taking half of the eigenfunctions, around the 
band center, of each random matrix. 
Each average is computed with $5\times 10^5$ data values.

\begin{figure}[t]
\centerline{\includegraphics[width=6cm]{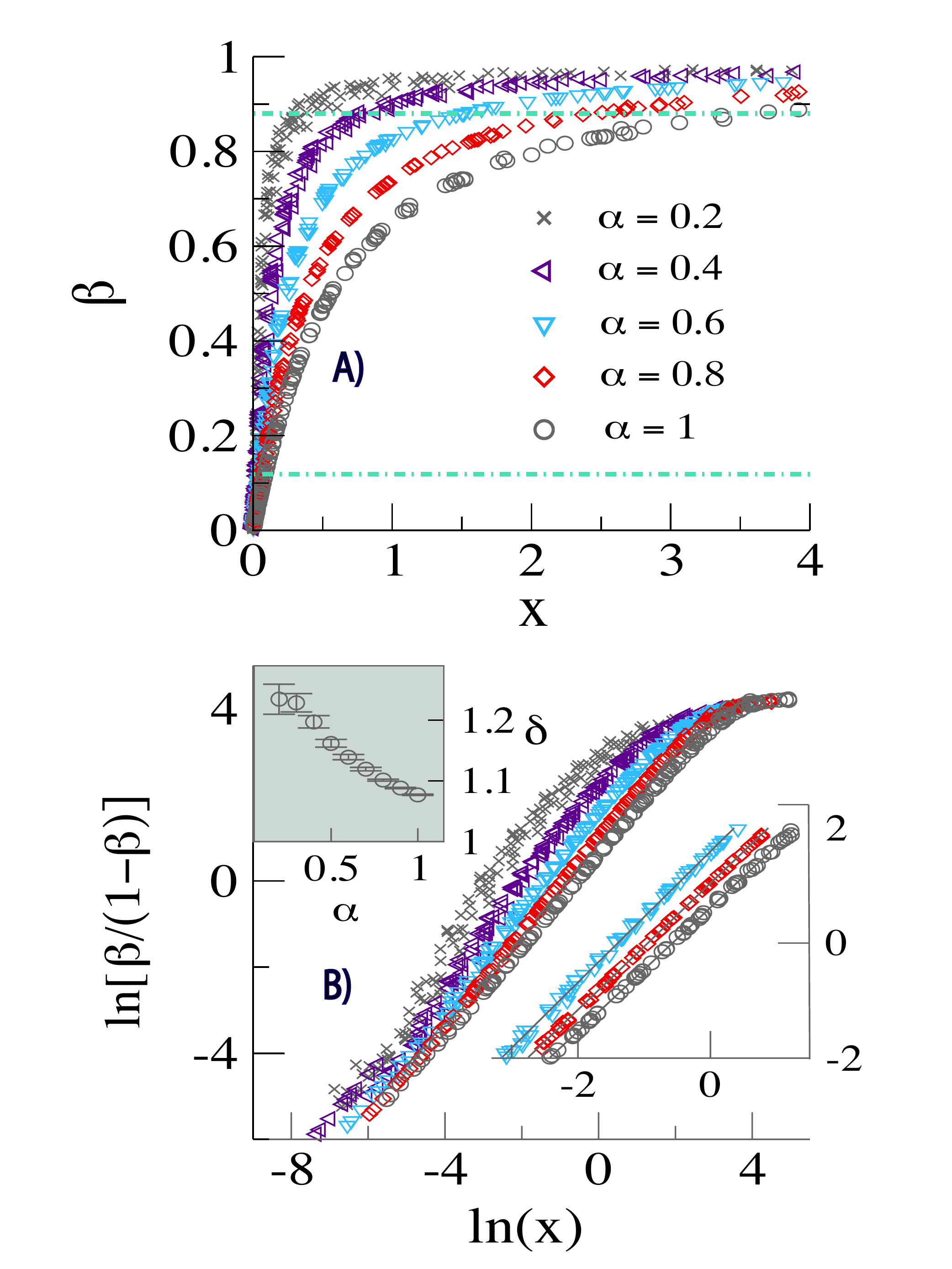}}
\caption{(Color online) 
(a) Scaled localization length $\beta$ as a function of $x=b_{\mbox{\tiny eff}}^2/N$ 
[see Eq.~(\ref{x})] for ensembles of diluted banded random matrices characterized by 
the sparsity $\alpha$. Horizontal black dot-dashed lines at 
$\beta\approx 0.12$ and 0.88 are shown as a reference, see the text.
(b) Logarithm of $\beta/(1-\beta)$ as a function of $\ln(x)$. 
Upper inset: Power $\delta$, from the fittings of the data with Eq.~(\ref{betascaling2}), 
as a function of $\alpha$. 
Lower inset: Enlargement in the range $\ln[\beta/(1-\beta)]=[-2,2]$ including data for 
$\alpha=0.6$, 0.8, and 1. Lines are fittings of the data with Eq.~(\ref{betascaling2}).}
\label{Fig1}
\end{figure}

In Fig.~\ref{Fig1}(a) we present $\beta$ as a function of $x$, see Eq.~(\ref{x}),
for ensembles of matrices characterized by the sparsity $\alpha$. 
We observe that the curves of $\beta$ vs.~$x$ have a functional 
form similar to that for the BRM model (corresponding to $\alpha=1$). 
In addition, in Fig.~\ref{Fig1}(b) the logarithm of 
$\beta/(1-\beta)$ as a function of $\ln(x)$ is presented. 
The quantity $\beta/(1-\beta)$ was useful in the study of the scaling properties of the BRM 
model \cite{CMI90,FM92} because $\beta/(1-\beta) = \gamma x$, which is equivalent to scaling
(\ref{betascaling0}), implies that a plot of $\ln[\beta/(1-\beta)]$  vs.~$\ln(x)$ is a straight line 
with unit slope. Even though, this statement is valid for the BRM model in a wide range of 
parameters (i.e.,~for $\ln[\beta/(1-\beta)]<2$) it does not apply to the dBRM model; see
Fig.~\ref{Fig1}(b). In fact, from this figure we observe that plots of 
$\ln[\beta/(1-\beta)]$ vs.~$\ln(x)$ are straight lines (in a wide range of $x$) with 
a slope that depends on the sparsity $\alpha$. Consequently, we propose the scaling law 
\begin{equation}
\frac{\beta}{1-\beta} = \gamma x^\delta \ ,
\label{betascaling2}
\end{equation}
where both $\gamma$ and $\delta$ depend on $\alpha$. Indeed, Eq.~(\ref{betascaling2})
describes well our data, mainly in the range $\ln[\beta/(1-\beta)]=[-2,2]$, as can be 
seen in the inset of Fig.~\ref{Fig1}(b) where we show the numerical data for 
$\alpha=0.6$, 0.8 and 1 and include fittings with Eq.~(\ref{betascaling2}).
We stress that the range $\ln[\beta/(1-\beta)]=[-2,2]$ corresponds to a
reasonable large range of $\beta$ values, $\beta\approx[0.12,0.88]$, whose bounds
are indicated with horizontal dot-dashed lines in Fig.~\ref{Fig1}(a).
Also, we notice that the power $\delta$, obtained from the fittings of the data 
using Eq.~(\ref{betascaling2}), is very close to unity for all the sparsity
values we consider here (see the upper inset of Fig.~\ref{Fig1}(b)).

\begin{figure}[t]
\centerline{\includegraphics[width=6cm]{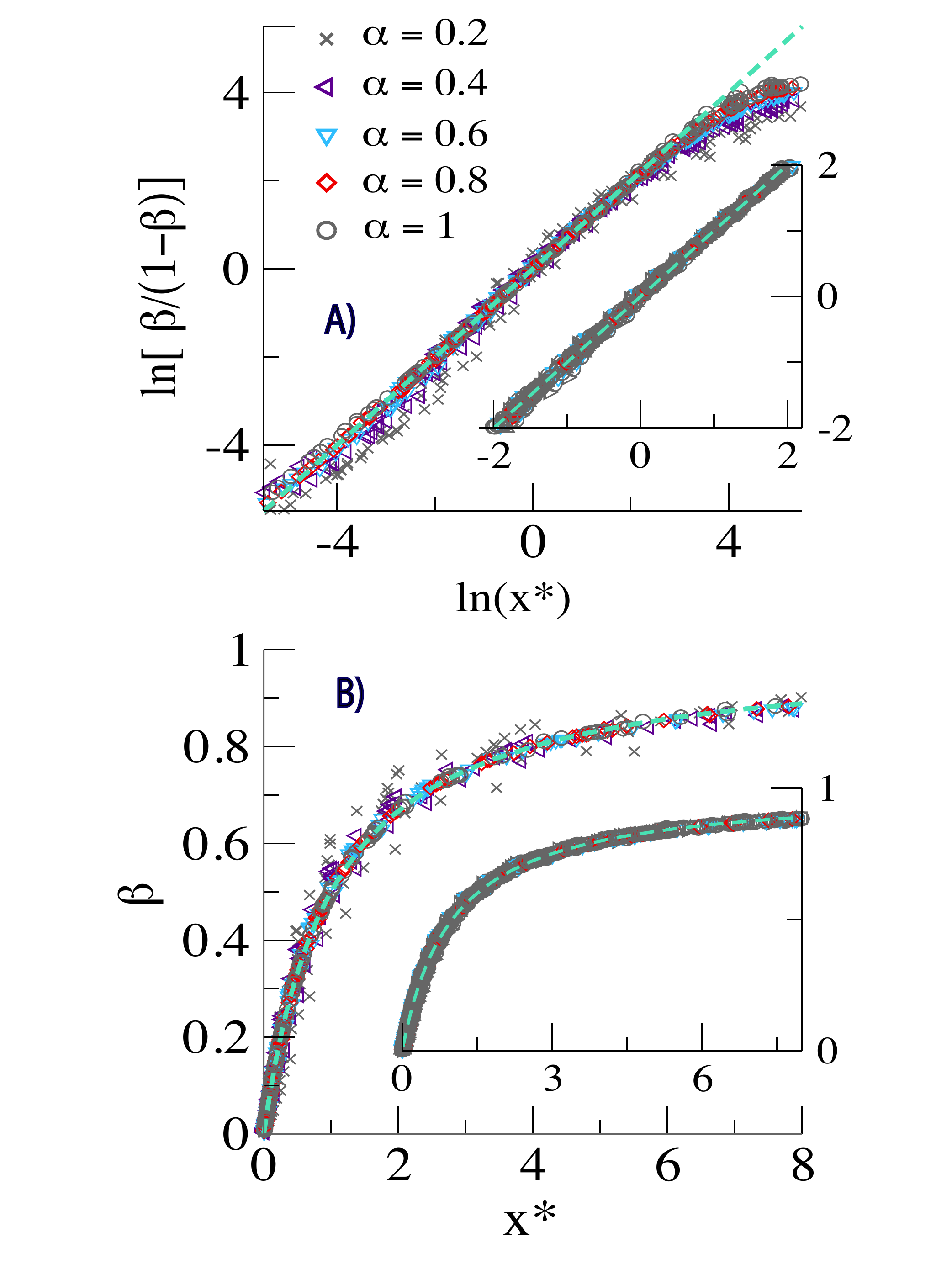}}
\caption{(Color online) 
(a) Logarithm of $\beta/(1-\beta)$ as a function of $\ln(x^*)$ [see Eq.~(\ref{betascaling3})]. 
Inset: Enlargement in the range $\ln[\beta/(1-\beta)]=[-2,2]$ including curves for 
$\alpha\in [0.5,1]$ in steps of $0.05$. Green dashed lines in main panel and inset 
are Eq.~(\ref{betascaling3}).
(b) $\beta$ as a function of $x^*$. 
Inset: Data for $\alpha\in [0.5,1]$ in steps of $0.05$. Green dashed lines in main 
panel and inset are Eq.~(\ref{betax*}).}
\label{Fig2}\end{figure}

Therefore, from the analysis of the data in Fig.~\ref{Fig1}, we are able to write down 
a {\it universal scaling function} for the scaled localization length $\beta$ of the
dBRM model as
\begin{equation}
\frac{\beta}{1-\beta} = x^* \ , \qquad x^*\equiv \gamma x^\delta \ .
\label{betascaling3}
\end{equation}
To validate Eq.~(\ref{betascaling3}) in Fig.~\ref{Fig2}(a) we present again the
data for $\ln[\beta/(1-\beta)]$ shown in Fig.~\ref{Fig1}(b) but now as a function 
of $\ln(x^*)$. We do observe that curves for different values of $\alpha$ fall on 
top of Eq.~(\ref{betascaling3}) for a wide range of the variable $x^*$.
Moreover, the collapse of the numerical data on top of Eq.~(\ref{betascaling3}) 
is excellent in the range $\ln[\beta/(1-\beta)]=[-2,2]$ for $\alpha\ge 0.5$, as
shown in the inset of Fig.~\ref{Fig2}(a).

Finally, we rewrite Eq.~(\ref{betascaling3}) into the equivalent, but explicit, 
scaling function for $\beta$:
\begin{equation}
\beta = \frac{x^*}{1+x^*} \ .
\label{betax*}
\end{equation}
In Fig.~\ref{Fig2}(b) we confirm the validity of Eq.~(\ref{betax*}).
We would like to emphasize that the universal scaling given in Eq.~(\ref{betax*})
extends outsize the range $\beta\approx[0.12,0.88]$, for which Eq.~(\ref{betascaling2})
was shown to be valid, see the main panel of Fig.~\ref{Fig2}(b). Furthermore, 
the collapse of the numerical data on top of Eq.~(\ref{betax*}) is remarkably good for 
$\alpha\ge 0.5$, as shown in the inset of Fig.~\ref{Fig2}(b).


{\it Spectral properties}.--
For completeness, now we analyze the spectral properties of the dBRM model. To this
end we choose $P(s)$, the {\it nearest neighbor energy-level spacing distribution}.
For $\alpha=0$ or $b=1$, i.e.~when the dBRM model produces diagonal matrices, $P(s)$ 
follows the exponential distribution $P(s) = \exp(-s)$; better known in RM theory as Poisson 
distribution or the spacing rule for random levels~\cite{Metha}.
In the opposite limit, for $\alpha=1$ and $b=N$, i.e.~when the dBRM reproduces the
GOE, $P(s)$ closely follows the Wigner-Dyson distribution~\cite{Metha}: 
$P(s) = (\pi/2) s \exp (-\pi s^2/4)$.
Then, by moving $\alpha$ and $b$ in the intervals $(0,1)$ and $(1,N)$, respectively, the 
$P(s)$ of the dBRM model should have a shape in-between the Poisson and Wigner-Dyson 
distributions. 

Here, in order to characterize the $P(s)$ for our RM model we use the 
phenomenological expression known as Izrailev's distribution~\cite{CIM91,I93}:
\begin{equation}
\label{izr}
P(s) = B_1z^{\widetilde{\beta}}(1+B_2\widetilde{\beta} z)^{f(\widetilde{\beta})} \exp\left[-\frac{1}{4}\widetilde{\beta} z^2-\left(1-\frac{\widetilde{\beta}}{2}\right)z \right] \ ,
\end{equation}
where $z=\pi s/2$,  $f(\widetilde{\beta})=\widetilde{\beta}^{-1}2^{\widetilde{\beta}}(1-\widetilde{\beta}/2)-0.16874$, and the parameters
$B_{1,2}$ are determined by the normalization conditions
$\int_0^\infty P(s)ds = \int_0^\infty sP(s)ds = 1$. We call $\widetilde{\beta}$ the spectral
parameter. 
In fact, Eq.~(\ref{izr}) has been shown to be useful to characterize the $P(s)$ of the BRM 
model~\cite{CIM91}, so we expect Eq.~(\ref{izr}) with $\widetilde{\beta}\in[0,1]$ to properly 
describe the $P(s)$ of the dBRM model. 

Thus, we construct histograms of  $P(s)$ 
for a large number of combinations of the parameters of the dBRM model ($\alpha ,b,N$) and 
by fitting them with Eq.~(\ref{izr}) we extract systematically the corresponding values of 
$\widetilde{\beta}$. We always construct $P(s)$ from half of the total unfolded~\cite{Metha} 
spacings $s_m=(E^{m+1}-E^m)/\Delta$ around the band center, where the density 
of states is approximately constant. Here, $E^m$ is the $m$-th eigenvalue and $\Delta$ the 
mean level spacing. Each histogram is constructed with $5\times 10^5$ spacings.

\begin{figure}[t]
\centerline{\includegraphics[width=6.5cm]{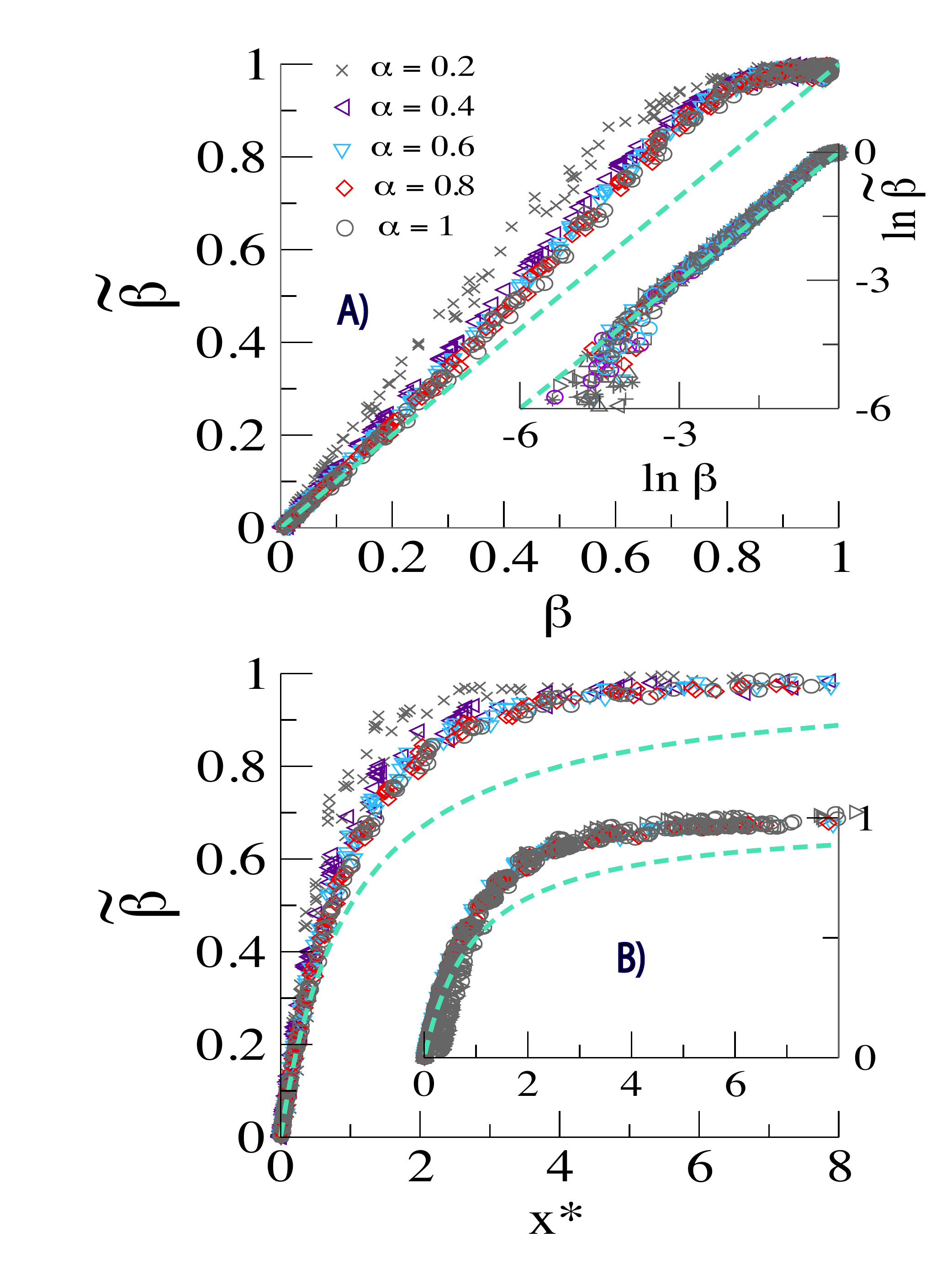}}
\caption{(Color online) 
(a) Spectral parameter $\widetilde{\beta}$  [see Eq.~(\ref{izr})] as a function of the 
scaled localization length $\beta$ for ensembles of diluted banded random matrices 
characterized by the sparsity 
$\alpha$. Inset: $\ln \widetilde{\beta}$ vs.~$\ln\beta$ for $\alpha\in [0.5,1]$ in steps of 
$0.05$. Green dashed lines in main panel and inset are the identity.
(b) Repulsion parameter $\widetilde{\beta}$ as a function of $x^*$. 
Inset: Data for $\alpha\in [0.5,1]$ in steps of $0.05$. 
Green dashed lines in main panel and inset are Eq.~(\ref{betax*}).}
\label{Fig3}
\end{figure}

In Fig.~\ref{Fig3}(a) we present the spectral parameter $\widetilde{\beta}$ as a function of the 
scaled localization length $\beta$ for the dBRM model. As in Figs.~\ref{Fig1} and \ref{Fig2}, here
we label different sparsities $\alpha$ with different colors (symbols). It is interesting to note that 
even though the relation between $\widetilde{\beta}$ and $\beta$ is not simple, e.g.~linear as
reported for other disordered systems~\cite{I89,I90,CCGI93,SIZC12,FGM13}, the curves 
$\widetilde{\beta}$ vs.~$\beta$ are independent of $\alpha$ once $\alpha>0.4$, see
inset of Fig.~\ref{Fig3}(a). This allows us to guess that $x^*$ can also serve to scale
the spectral parameter $\widetilde{\beta}$, at least for $\alpha>0.4$.
Accordingly, in Fig.~\ref{Fig3}(b) we show that the curves of $\widetilde{\beta}$ vs.~$x^*$ 
fall one on top of the other mainly for $\alpha>0.4$, see the inset of the figure.

From Fig.~\ref{Fig3}(b) we also observe that the curves $\widetilde{\beta}$ vs.~$x^*$ are
above Eq.~(\ref{betax*}), that we include as dashed lines, except for very small values of $x^*$
where they coincide. This fact has already been reported for the BRM model in Ref.~\cite{I95}.
This also means that the spectral properties of the dBRM model approach the GOE limit faster 
than the eigenfunction properties.

{\it Conclusions}.--
In this paper, by using extensive numerical simulations, we demonstrate that the 
normalized localization length $\beta$ of the eigenfunctions of a diluted banded random 
matrix (dBRM) model
scales with the parameter $x^*(N,b,\alpha)=\gamma(\alpha)[(b\alpha)^2/N]^{\delta(\alpha)}$ 
as $x^*/(1+x^*)$, where $(N,b,\alpha)$ are the model parameters (matrix size, bandwidth, 
and sparsity, respectively) and $\gamma$ and $\delta$ are scaling parameters.
In addition, by plotting the spectral parameter $\widetilde{\beta}$ (the repulsion parameter of
Izrailev's distribution) as a function of $\beta$ we realized that, for moderate sparsity 
(i.e.~$\alpha>0.4$), $x^*(N,b,\alpha)$ also scales the spectral properties of the dBRM model.

While general diluted RM models have direct applications to random networks 
(i.e.~the adjacency matrices of complex networks are, in general, diluted random 
matrices), the dBRM 
model may be used to model multilayer random networks since the bandwidth 
$b$ and the sparsity $\alpha$ can be associated, respectively, to the size and 
connectivity of the subnetworks composing a multilayer, see e.g.~\cite{MFRM16}.

Finally, we want to recall that the scaling (\ref{betascaling0}), valid for the BRM model
(and other disordered systems~\cite{CGIS90,CGIFM92}),
was rewritten in a more elegant way as a relation between properly-defined inverse 
lengths~\cite{FM92,CGIFM92}
\begin{equation}
\label{1/d}
[d(N,W)]^{-1} = [d(\infty,W)]^{-1} + [d(N,0)]^{-1}  \ ,
\end{equation}
were $d(N,W)\equiv \exp[\bra S(N,W) \ket]$ and $W$ represents $b$ for the BRM 
model or the localization length for the one-dimensional Anderson model
and Lloyd's model.
Here, in the case of the dBRM model, scaling (\ref{betax*}) can also be written in
the ``model independent" form (\ref{1/d}) as
\begin{equation}
\label{1/deff}
[d(N,b,\alpha)]^{-1} = [d(\infty,b,\alpha)]^{-1} + [d(N,N,1)]^{-1} \ ,
\end{equation}
with $d(N,b,\alpha)\equiv\exp[\bra S(N,b,\alpha) \ket]$ and 
$d(N,N,1)=\exp[S_{\tbox{GOE}}(N)]$ (the reference entropy).

We hope our results may motivate a theoretical approach to the dBRM model.

{\it Acknowledgements}.--
This work was partially supported by VIEP-BUAP (Grant No.~MEBJ-EXC17-I), 
Fondo Institucional PIFCA (Grant No.~BUAP-CA-169), and 
CONACyT (Grant No.~CB-2013/220624).
FAR acknowledges CNPq (Grant No.~305940/2010-4), 
FAPESP (Grant No.~2011/50761-2 and 2013/26416-9), and 
NAP eScience - PRP - USP for financial support. 
GFA would like to acknowledge FAPESP (grants 2012/25219-2 and 2015/07463-1) 
for the scholarship provided. Y. M. acknowledges support from the Government of Arag\'on, Spain through a grant to the group FENOL, and by MINECO and FEDER funds (grant FIS2014-55867-P).

\bibliographystyle{plainnat}

\begin{thebibliography}{99}


\bibitem{Metha}
M. L. Metha,
{\it Random matrices} (Elsevier, Amsterdam, 2004).

\bibitem{RMT}
{\it The Oxford Handbook of Random Matrix Theory}, 
G. Akemann, J. Baik, and P. Di Francesco (Eds.)
(Oxford University Press, New York, 2011).

\bibitem{KKCA}
V. E. Kravtsov, I. M. Khaymovich, E. Cuevas, and M. Amini,
A random matrix model with localization and ergodic transitions,
New J. Phys. {\bf 17}, 122002 (2015).

\bibitem{W55}
E. P. Wigner,
Characteristic vectors of bordered matrices with infinite dimensions,
Ann. Math. {\bf 62}, 548 (1955).

\bibitem{W57}
E. P. Wigner,
Ann. Math. {\bf 65}, 203 (1957); SIAM Review {\bf 9}, 1 (1967).


\bibitem{WFL91}
M. Wilkinson, M. Feingold, and D. M. Leitner,
Localization and spectral statistics in a banded random matrix ensemble,
J. Phys. A: Math. Gen. {\bf 24}, 175 (1991).

\bibitem{FLW91}
M. Wilkinson, M. Feingold, and D. M. Leitner,
Spectral statistics in semiclasical random-matrix ensembles,
Phys. Rev. Lett. {\bf 66}, 986 (1991).

\bibitem{FGIM93}
M. Feingold, A. Gioletta, F. M. Izrailev, and L. Molinari,
Two parameter scaling in the Wigner ensemble,
Phys. Rev. Lett. {\bf 70}, 2936 (1993).

\bibitem{CCGI93}
G. Casati, B. V. Chirikov, I. Guarneri, and F. M. Izrailev,
Band-random-matrix model for quantum localization in conservative systems,
Phys. Rev. E {\bf 48}, R1613 (1993).

\bibitem{CCGI96}
G. Casati, B.V. Chirikov, I. Guarneri, and F. M. Izrailev,
Quantum ergodicity and localization in conservative systems:
the Wigner band random matrix model,
Phys. Lett. A {\bf 223}, 430 (1996).

\bibitem{F97}
M. Feingold,
Localization in strongly chaotic systems,
J. Phys. A: Math. Gen. {\bf 30}, 3603 (1997).

\bibitem{W00}
W. Wang,
Perturbative and nonperturbative parts of eigenstates and local spectral density of states:
The Wigner-band random-matrix model,
Phys. Rev. E {\bf 61}, 952 (2000).

\bibitem{W01}
W. Wang,
Approach to energy eigenvalues and eigenfunctions from nonperturbative regions
of eigenfunctions,
Phys. Rev. E {\bf 63}, 036215 (2001).


\bibitem{MFDQS96}
A. D. Mirlin, Y. V. Fyodorov, F.-M. Dittes, J. Quezada, and T. H. Seligman, 
Transition from localized to extended eigenstates in the ensemble of power-law 
random banded matrices,
Phys. Rev. E {\bf 54}, 3221 (1996).

\bibitem{EM08}
F. Evers and A. D. Mirlin, 
Anderson transitions,
Rev. Mod. Phys. {\bf 80}, 1355 (2008).


\bibitem{CMI90}
G. Casati, L. Molinari, and F. M. Izrailev,
Scaling properties of band random matrices,
Phys. Rev. Lett. {\bf 64}, 1851 (1990).

\bibitem{EE90}
S. N. Evangelou and E. N. Economou,
Eigenvector statistics and multifractal scaling of band random matrices,
Phys. Lett. A {\bf 151}, 345 (1990).

\bibitem{FM91}
Y. F. Fyodorov and A. D. Mirlin,
Scaling properties of localization in random band matrices: A $\sigma$-model approach,
Phys. Rev. Lett. {\bf 67}, 2405 (1991).

\bibitem{BMP91}
L. V. Bogachev, S. A. Molchanov, L. A. Pastur,
On the level density of random band matrices,
Mathematical Notes {\bf 50:6}, 1232 (1991).

\bibitem{MPK92}
S. A. Molchanov, L. A. Pastur, A. M. Khorunzhii,
Limiting eigenvalue distribution for band random matrices,
Theor. Math. Phys. {\bf 90:2}, 108 (1992).

\bibitem{I95}
F. M. Izrailev,
Scaling properties of spectra and eigenfunctions for quantum dynamical and 
disordered systems,
Chaos Solitons Fractals {\bf 5}, 1219 (1995).

\bibitem{FM92}
Y. F. Fyodorov and A. D. Mirlin,
Analytical derivation of the scaling law for the inverse participation ratio in 
quasi-one-dimensional disordered systems,
Phys. Rev. Lett. {\bf 69}, 1093 (1992).

\bibitem{MF93}
A. D. Mirlin and Y. F. Fyodorov,
The statistics of eigenvector components of random band matrices: Analytical results,
J. Phys. A: Math. Gen. {\bf 26}, L551 (1993).

\bibitem{FM93}
Y. F. Fyodorov and A. D. Mirlin,
Level-to-level fluctuations of the inverse participation ratio in finite quasi 
1D disordered systems,
Phys. Rev. Lett. {\bf 71}, 412 (1993).

\bibitem{FM94}
Y. F. Fyodorov and A. D. Mirlin,
Statistical properties of eigenfunctions of random quasi 1D one-particle
Hamiltonians,
Int. J. Mod. Phys. B {\bf 8}, 3795 (1994).

\bibitem{CIM91}
G. Casati, F. M. Izrailev, and L. Molinari,
Scaling properties of the eigenvalue spacing distribution for band random matrices,
J. Phys. A: Math. Gen. {\bf 24}, 4755 (1991).

\bibitem{MF96}
T. Kottos, A. Politi, F. M. Izrailev, and S. Ruffo,
Scaling properties of Lyapunov spectra for the band random matrix model,
Phys. Rev. E {\bf 53}, R5553 (1996).

\bibitem{CGM97}
G. Casati, I. Guarneri, and G. Maspero,
Landauer and Thouless conductance: A band random matrix approach,
J. Phys. I (France) {\bf 7}, 729 (1997).

\bibitem{KPI98}
T. Kottos, A. Politi, and F. M. Izrailev,
Finite-size corrections to Lyapunov spectra for band random matrices,
J. Phys.: Condens. Matter {\bf 10}, 5965 (1998).

\bibitem{KIP99}
T. Kottos, F. M. Izrailev, and A. Politi,
Finite-length Lyapunov exponents and conductance for quasi-1D disordered solids,
Physica D {\bf 131}, 155 (1999).

\bibitem{P01}
P. Shukla,
Eigenvalue correlations for banded matrices,
Physica E {\bf 9}, 548 (2001).

\bibitem{W02}
W. Wang,
Localization in band random matrix models with and without increasing diagonal elements,
Phys. Rev. E {\bf 65}, 066207 (2002).


\bibitem{MF75}
K. K. Mon and J. B. French, 
Statistical properties of many-particle spectra,
Ann. Phys. (N.Y.) {\bf 95}, 90 (1975).

\bibitem{BW03}
L. Benet and H. A. Weidenm\"uller,
Review of the $k$-body embedded ensembles of Gaussian random matrices,
J. Phys. A: Math. Gen.  {\bf 36}, 3569 (2003).

\bibitem{K14}
V. K. B. Kota,
{\it Embedded random matrix ensembles in quantum physics},
Lecture Notes in Physics 884 (Springer, London, 2014).
 

\bibitem{CK01}
D. Cohen and T. Kottos, 
Parametric dependent Hamiltonians, wave functions, random matrix theory, 
and quantal-classical correspondence,
Phys. Rev. E {\bf 63}, 036203 (2001).

\bibitem{CH00}
D. Cohen and E. J. Heller, 
Unification of perturbation theory, random matrix theory, and semiclassical 
considerations in the study of parametrically dependent eigenstates,
Phys. Rev. Lett. {\bf 84}, 2841 (2000).

\bibitem{FP86}
M. Feingold and A. Peres,
Distribution of matrix elements of chaotic systems,
Phys. Rev. A {\bf 34}, 591 (1986).


\bibitem{S94}
D. L. Shepelyansky,
Coherent propagation of two interacting particles in a random potential,
Phys. Rev. Lett. {\bf 73}, 2607 (1994).

\bibitem{FM95a}
Y. V. Fyodorov and A. D. Mirlin,
Statistical properties of random banded matrices with strongly fluctuating diagonal elements,
Phys. Rev. B {\bf 52}, R11580 (1995).

\bibitem{FM95b}
Y. V. Fyodorov and A. D. Mirlin,
Analytical results for random band matrices with preferential basis,
Europhys. Lett., {\bf 32}, 385 (1995).

\bibitem{S97}
P. G. Silvestrov,
Summing graphs for random band matrices,
Phys. Rev. E {\bf 55}, 6419 (1997).

\bibitem{DPS02}
M. Disertori, H. Pinson, and T. Spencer,
Density of states for random band matrices,
Commun. Math. Phys. {\bf 232}, 83 (2002).

\bibitem{KK02}
A. Khorunzhy and W. Kirsch,
On asymptotic expansions and scales of spectral universality in band 
random matrix ensembles,
Commun. Math. Phys. {\bf 231}, 223 (2002).

\bibitem{S09}
J. Schenker,
Eigenvector localization for random band matrices with power law band width,
Commun. Math. Phys. {\bf 290}, 1065 (2009).

\bibitem{S10}
S. Sodin, 
The spectral edge of some random band matrices,
Annals of Math. {\bf 172}, 2223 (2010).


\bibitem{FCIC96}
Y. V. Fyodorov, O. A. Chubykalo, F. M Izrailev, and G. Casati,
Wigner random banded matrices with sparse structure: Local spectral density of states,
Phys. Rev. Lett. {\bf 76}, 1603 (1996).


\bibitem{CRBD16}
X. Cao, A. Rosso, J.-P. Bouchaud, P. LeDoussal,
Genuine localisation transition in a long-range hopping model,
arXiv:1607.04173.

\bibitem{MRV16}
J. A. Mendez-Bermudez, F. A. Rodrigues, and D. A. Vega-Oliveros,
Multifractality in random networks with long-range spatial correlations.
to be submitted.

\bibitem{MFRM16}
J. A. Mendez-Bermudez, G. Ferraz-de-Arruda, F. A. Rodrigues, and Y. Moreno,
Scaling properties of multilayer random networks,
arXiv:1611.06695.


\bibitem{RB88}
G. J. Rodgers and A. J. Bray,
Density of states of a sparse random matrix,
Phys. Rev. B {\bf 37}, 3557 (1988).

\bibitem{FM91b}
Y. V. Fyodorov and A D Mirlin,
On the density of states of sparse random matrices,
J. Phys. A: Math. Gen. {\bf 24}, 2219 (1991).

\bibitem{MF91}
A. D. Mirlin and Y. V. Fyodorov,
Universality of level correlation function of sparse random matrices,
J. Phys. A: Math. Gen. {\bf 24}, 2273 (1991).

\bibitem{E92}
S. N. Evangelou,  
A numerical study of sparse random matrices,
J. Stat. Phys. {\bf 69}, 361 (1992).

\bibitem{JMR01}
A. D. Jackson, C. Mejia-Monasterio, T. Rupp, M. Saltzer, and T. Wilke,
Spectral ergodicity and normal modes in ensembles of sparse matrices,
Nucl. Phys. A {\bf 687}, 405 (2001).

\bibitem{KSV04}
O. Khorunzhy, M. Shcherbina, and V. Vengerovsky,
Eigenvalue distribution of large weighted random graphs,
J. Math. Phys. {\bf 45}, 1648 (2004).

\bibitem{K08}
R. K\"uhn,
Spectra of sparse random matrices,
J. Phys. A: Math. Theor. {\bf 41}, 295002 (2008).

\bibitem{S09b}
S. Sodin, 
The Tracy-Widom law for some sparse random matrices,
J. Stat. Phys. {\bf 136}, 834 (2009).

\bibitem{SC012}
G. Semerjian and L. F. Cugliandolo,
Sparse random matrices: the eigenvalue spectrum revisited,
J. Phys. A: Math. Gen. {\bf 35}, 4837 (2002).

\bibitem{EKYY13}
L. Erd\"os, A. Knowles, H.-T. Yau, and J. Yin,
Spectral statistics of Erd\"os-R\'enyi graphs I: Local semicircle law,
Ann. Probab. {\bf 41}, 2279 (2013).

\bibitem{EKYY12}
L. Erd\"os, A. Knowles, H.-T. Yau, and J. Yin,
Spectral statistics of Erd\"os-R\'enyi graphs II: Eigenvalue spacing and the extreme eigenvalues,
Commun. Math. Phys. {\bf 314}, 587 (2012).

\bibitem{MAMRP15}
J. A. Mendez-Bermudez, A. Alcazar-Lopez, A. J. Martinez-Mendoza, 
F. A. Rodrigues, and T. K. DM. Peron,
Universality in the spectral and eigenfunction properties of random networks,
Phys. Rev. E {\bf 91}, 032122 (2015).



\bibitem{CGIS90}
G. Casati, I. Guarneri, F. M. Izrailev, and R. Scharf, 
Scaling behavior of localization in quantum chaos, 
Phys. Rev. Lett. {\bf 64}, 5 (1990).

\bibitem{I90}
F. M. Izrailev,
Simple models of quantum chaos: Spectrum and eigenfunctions,
Phys. Rep. {\bf 196}, 299 (1990).


\bibitem{CGIFM92}
G. Casati, I. Guarneri, F. M. Izrailev, S. Fishman, and L. Molinari, 
Scaling of the information length in 1D tight-binding models, 
J. Phys.: Condens. Matter {\bf 4}, 149 (1992).


\bibitem{I93}
F. M. Izrailev, 
in Quantum Chaos, Proceedings of the International
School of Physics ``Enrico Fermi'', Course CXIX, Varenna,
1991, edited by G. Casati, I. Guarneri, and U. Smilansky
(North-Holland, Amsterdam, 1993) p. 265.

\bibitem{I89}
F. M. Izrailev, 
Intermediate statistics of the quasi-energy spectrum and quantum localisation of classical chaos,
J. Phys. A {\bf 22}, 865 (1989).

\bibitem{SIZC12}
S. Sorathia, F. M. Izrailev, V. G. Zelevinsky, and G. L. Celardo,
From closed to open one-dimensional Anderson model: Transport versus spectral statistics,
Phys. Rev. E {\bf 86}, 011142 (2012).

\bibitem{FGM13}
J. Flores, L. Gutierrez, R. A. Mendez-Sanchez, G. Monsivais, P. Mora, and A. Morales, 
Anderson localization in finite disordered vibrating rods,
Europhys. Lett. {\bf 101}, 67002 (2013).




      

\end{thebibliography}

\end{document}